\documentclass{article}

\title{Planetary interior and habitability of exoplanets: Recent developments}
\author{Nisha Katyal}
\date{April 2021}

\usepackage{natbib}
\usepackage{graphicx}
\usepackage{chemformula}
\usepackage[breaklinks,colorlinks,citecolor=blue]{hyperref} 
\begin{document}

\maketitle

\section{Introduction}

Out of the various definitions of habitability commonly used in the field of astrobiology \citep{Abel2020}, the most canonical definition of habitability used for the exoplanetary science is based on the habitable zone i.e the presence of surface liquid water on a planet \citep{Kasting1993}. It is important to understand the origin of water on a planetary surface because it is essential for determining the planetary habitability and whether the planet can support life. With the discoveries of new extra-solar planets over last decade, significant work has been dedicated to understanding their habitability and whether there is a possibility of life on these planets as determined by the presence of biosignatures produced through biotic or abiotic pathways \citep{Joshua2021,Rita2021,Hasan2021}.

For terrestrial solar system planets, early development and evolution of a planetary atmosphere after its formation e.g. during the hot magma ocean phase on early Earth \citep{Katyal2019} has resulted in large oceans of liquid water- a requirement for habitability of a planet. It is intriguing to understand whether such hot conditions exist/existed on the extrasolar planets and what is their thermal cooling mechanism in order to achieve habitable conditions.  An interesting study by \cite{Godolt2019} explored the  
potential of secondary outgassing from the planetary interior to rebuild a water reservoir allowing for habitability at a later stage. They arrived at an important conclusion that Earth-like stagnant lid planets could allow for secondary atmospheres even after severe water loss or re-gassing (sequestering) of \ch{H2O} in the mantle to obtain habitable surface conditions within a continuous habitable zone. In this context, it is important to connect the interiors specially its composition to the atmosphere and study the coupled evolution of the number of rocky exoplanets discovered so far.

The interior dynamics of a planet not only affects its climate but also controls the long-term evolution of important greenhouse gases such as \ch{CO2} which in turn affects the habitability of the planet \citep{Oosterloo2021}. In fact, the outer edge of the habitable zone strongly depends on the amount of \ch{CO2} outgassed from the interior \citep{Godolt2019,Graham2021}. Along with that, other  parameters that characterize the atmosphere of exoplanets are the size, radius of a planet and distribution of mass in its core that play an important role in determining the outgassing of volatile species into the atmosphere (a positive feedback). On the other hand, the mechanism of silicate weathering is responsible for removing \ch{CO2} (negative feedback) from the atmosphere back to the seafloor which ultimately effects the long term stability of the climate over geological timescales. Hence, in order to study the pathway to habitability of a planet, its important to take into consideration various positive and negative feedback mechanisms between the interior, surface and atmosphere.

 This article deals with the most recent developments in the field of exoplanetary science connecting the interior of the planets with their habitability. In this issue, I have specified the importance of interior dynamics and briefly reviewed some of the main factors by which interior of a planet can effect the habitability of extra-solar planets.

\section{Earth-sized rocky planets: Hot to habitable conditions}
A hot and uninhabited planet may take several thousand years to cool down to have early habitable conditions or it may remain in the same state. One of the important factors responsible for planetary cooling is the  composition of the atmosphere. A recent study by \cite{Tim2021} analysed the outgassing sequence of various volatiles from the interior and studied the thermal evolution of the atmospheres in a coupled interior-atmospheric modeling framework. They concluded that an atmosphere with gases such as CO, \ch{N2}, \ch{O2} will cool down quickly, \ch{H2O}, \ch{CO2}, \ch{CH4} will cool down intermediately and  a thick \ch{H2} atmosphere will take the longest to cool down. This is because of the emission properties of the gases in the atmosphere as discussed in their study. Another study by \cite{Katyal2020} studied the atmospheric evolution of early Earth during the magma ocean phase with hot surface conditions and how the surface cools down to allow water to exist in liquid state. It is, however, not clear how the steam condensed to became large water oceans from which the life is understood to have originated on Earth and hence, needs further investigation beyond the magma ocean phase. 

\subsection{Sequestering of \ch{O2} back to mantle}
Enrichment of planetary interior or regassing can happen due to photo-dissociation of \ch{H2O} in the atmosphere by which the oxygen can get sequestered back to the mantle, hence increasing the mantle oxidation state \citep{Ortenzi2020}. Although, there is a possibility that \ch{O2} stays in the atmosphere for some of the highly irradiated planets of TRAPPIST-1 planetary system  \citep{Barth2020}, it is unknown for how long these planets can sustain oxygen atmosphere provided the strong XUV radiation from the star can cause hydrodynamic loss of the molecule. Tendency to lose an atmosphere via such processes could be tested by transit observations as attempted recently for the rocky exoplanet LHS 3844b orbiting a M-star where no atmosphere was being found \citep{Diamond2020}. 

\subsection{\ch{O2} in the atmosphere as false positives}
On the other hand, a thin \ch{O2} atmosphere originated with this process is a false positive and could be misunderstood as a potential biosignature. A recent study by \cite{Joshua2021} have explored possible formation scenarios of non-biological oxygen in the atmosphere for a variety of initial volatile inventories and categorized them as waterworlds, desert-worlds, high \ch{CO2}:\ch{H2O} perpetual runaway greenhouse. If the initial inventories of planets orbiting G-stars are Earth-like, its not possible to accumulate \ch{O2}-rich atmospheres. However, for the cases where volume fraction of \ch{CO2} is greater than volume fraction of \ch{H2O} $f_{\ch{CO2}} > f_{\ch{H2O}}$, mantle no longer acts as a oxygen sink, and (2) $f_{\ch{CO2}} > 1$, the crustal production is halted after a few billion years and shuts off all the oxygen sinks \citep{Joshua2021}. 

\subsection{Dependence on CMF and WMF}
Planetary cooling rate estimation based on the outgoing longwave radiation (OLR) emitted by the planets as a function of water mass fraction (WMF) and core mass fraction (CMF) in the mantle determine whether water exists as steam or liquid/ice in the atmosphere. An interior model with an MCMC Bayesian algorithm of \cite{Dorn2018} was used to estimate the posterior distributions of the core and water compositional parameters i.e. WMF and CMF of TRAPPIST-1 planets by a recent study by \cite{Lorena2021}. According to this study, planets TRAPPIST1-b, c are highly irradiated with large steam atmospheres and are in a runaway greenhouse state (absorbed radiation is higher than the OLR) These steam hydrospheres (densities) are comparable to dry rocky planets with no atmosphere. On the other hand, the hydrospheres of planet d to g are denser because it consists of condensed high-pressure ices and a higher WMF as compared to the TRAPPIST1-b,c that are close-in to the star. The maximum value of WMF is obtained for planet TRAPPIST1-d and estimated to be $\sim$ 20\% in this study. Moreover, for this planet, absorbed radiation is smaller than the emitted OLR. Hence, this planet can cool enough to maintain its atmosphere in condensed phase \citep{Lorena2021} and be a potentially habitable planet.

\section{Assumptions beyond considering Earth-sized planets}
  Previously reported habitable super-Earth K2-18b \citep{Benneke2019,Tsiaras2019}, which is twice the size of Earth and 8 times massive than Earth was further confirmed to be a potential candidate for habitability by a recent study of \citep{Nikku2020}. The latter study obtained a water vapour mixing ratio of 0.02-14.80\% in the atmosphere. According to these studies, K2-18b is so far the best potential candidate in the search of habitable planets. Additionally, because of their big size, there is a higher probability of their detection via the various current and future space missions such as James Web Space telescope (JWST) to be launched in 2021. Below, I discuss some of the factors on which habitability of super-Earths might depend upon.

\subsection{Dependence on CMF and WMF}
A wide range of core mass fractions (CMF) and water mass fractions (WMF) can be set as constraints while modeling the interior of super-Earths depending upon their mass and radius. By studying the variation of these parameters in the interior of K2-18b, \citep{Nikku2020} obtained the mass fraction of H/He envelope less than and equivalent to 10$^{-3}$ in the atmosphere and a reasonable CMF $<$15\% that allows for liquid water at Earth-like habitable conditions. For higher CMFs, H/He mass fractions are estimated to be higher that do not define the \ch{H2O}-H/He boundary (HBB) properly e.g. water is in super-critical phase. Hence, to determine a planet's habitability, it's important to obtain the extent of H/He envelope through a careful investigation of planetary interior and obtaining the thermodynamic conditions at HBB. Future observations with JWST will put more light on these findings.

\subsection{Dependence on mass of the planet}
A planet massive than Earth has a higher surface gravity and tends to have higher lithospheric pressure of various outgassed volatiles according to the relation M$_{\rm volatile}$ = P$\times$ A/g; where $P$ is the total surface pressure, $A$ is the area of the planet and $g$ is the gravitational acceleration. The partial pressure of \ch{CO2} could vary from 10-10$^{3}$ Pa for planetary mass 0.1-2 M$_{\rm E}$ at about 10 Gyr since its planetary evolution began \citep{Oosterloo2021}. The increased pressure at the surface leads to a higher melting temperature for planetary masses higher than Earth \citep{Oosterloo2021, Ortenzi2020}. The latter study showed that as a result of delay in production of melt, there could be very little outgassing or no outgassing. Hence, outgassing by a planet is primarily dependent upon its mass and thermal state \citep{Dorn2018}.
Therefore, it is expected that the super-Earths more massive than 4M$_{\rm E}$ will not outgas any volatiles from their interiors.

\section{Surface feedback processes}
The process of removal of \ch{CO2} from the atmosphere occurs by sequestration of \ch{CO2} to the oceans via the process known as silicate weathering. This is understood to be a negative feedback mechanism that stabilizes the long-term climate over geological time scales. Signatures of silicate weathering feedbacks can be, in fact, detected indirectly by the level of \ch{CO2} present in the atmosphere.

\cite{Graham2021} have used climate-chemistry models to demonstrate how perturbations in p\ch{CO2} minimally effect the surface temperature and ocean pH on large p\ch{CO2} planets in the innermost region of the habitable zone. On the other hand, planets near the outer edge of liquid water habitable zone require $\sim 10^{6} \times$ more \ch{CO2} to maintain surface temperature for liquid water to exist owing to the logarithm radiative impact of p\ch{CO2}. Previously, \cite{Graham2020} presented an improved version of weathering to detect signatures of silicate weathering feedbacks through exoplanet observations.

\section{Conclusion}
In the quest to understand habitability of recently discovered exoplanets, a precise knowledge of the various factors affecting the climate over geological time scales is required. This issue focuses on determining the habitability of extra-solar planets through a precise understanding of various interior parameters that include (1) core mass fraction (CMF) and water mass fraction (WMF) in the planetary interior, (2) stellar insolation, (3) mass fraction of H/He envelope in the atmosphere, and (4) loss mechanisms such as photo-dissociation of \ch{H2O} and removal of \ch{CO2} from the atmosphere via silicate weathering. As a consequence, the partial pressure of p\ch{CO2} at the surface plays an important role in controlling the climate hence strongly impacts planetary habitability for the reason that it influences the \ch{H2O} outgassing to an extent. The silicate weathering is another major important mechanism for controlling p\ch{CO2} on rocky exoplanets and stabilizing its climate over long-time evolution. The influence of star on the habitability of a planet is recently discussed in detail by \cite{Turbet2021}. 

Finally, although it is challenging to include all the factors and processes discussed above in a coupled framework including interior-surface-weathering-atmospheres, efforts should be made to connect the different sub-fields together to present the big picture of habitability of exoplanets.

\bibliographystyle{aa}
\bibliography{shorttitles,references}
\end{document}